\documentclass[letterpaper, 10 pt, conference]{ieeeconf}  

\IEEEoverridecommandlockouts                              

\usepackage{graphics} 
\usepackage{graphicx}
\usepackage{amsmath} 

\usepackage{cite}

\title{\LARGE \bf
Earable Platform with Integrated Simultaneous EEG Sensing and Auditory Stimulation
}

\author{Min Suk Lee$^{1}$, Abhinav Uppal$^{1}$, Ananya Thota$^{1}$, Chetan Pathrabe$^{1}$, Rommani Mondal$^{1}$, \\Akshay Paul$^{2}$, Yuchen Xu$^{2,*}$ and Gert Cauwenberghs$^{1,2,*}$
\thanks{$^{1}$Shu Chien-Gene Lay Department of Bioengineering, UC San Diego, La Jolla, CA 92093, USA.} 
\thanks{$^{2}$Institute for Neural Computation, UC San Diego, La Jolla, CA 92093, USA.}
\thanks{$^{*}$Contact: yux013@ucsd.edu, gcauwenberghs@ucsd.edu}
\thanks{© 2025 IEEE. Personal use of this material is permitted. Permission from IEEE must be obtained for all other uses, in any current or future media, including reprinting/republishing this material for advertising or promotional purposes, creating new collective works, for resale or redistribution to servers or lists, or reuse of any copyrighted component of this work in other works.
This work has been accepted for publication in the Proceedings of the 2025 IEEE International Conference on Neural Engineering (NER 2025).}
}

\begin{document}

\maketitle
\thispagestyle{empty}
\pagestyle{empty}

\begin{abstract}
Conventional scalp-based EEG systems are cumbersome to use, requiring extensive setup, restrictive wiring, and conductive gels that can dry out and limit long-term monitoring, while also carrying social stigma. As a result, there is increasing interest in in-ear EEG technology to improve comfort, convenience, and discretion for users. This work presents a personalized in-ear EEG monitor (IEEM) that simultaneously captures EEG signals from the outer ear while delivering audio playback through the same device. The earpiece is custom-molded to precisely match the user's ear anatomy, providing effective sound isolation from the environment and enabling direct audio transmission into the ear canal. Testing of the assembled earpiece shows successful detection of electrooculography (EOG), eye blinks, jaw clenches, auditory steady-state responses (ASSR), and alpha modulation. Electrochemical impedance spectroscopy (EIS) measurements confirm stable electrode-skin contact, with impedance values similar to those of traditional dry electrodes. The integrated approach enables potential closed-loop neuromodulation applications all in the ear where brain activity can be monitored in real-time and corresponding acoustic stimulation delivered adaptively.

\end{abstract}

\section{Introduction}

\indent There is an increasing demand in both the industry and research sphere for reliable, long-term monitoring of brain activity in users’ everyday lives. However, traditionally used EEG and brainwave reading devices are uncomfortable, obstructive, and simply inconvenient to wear in one's daily life. Additionally, they are socially stigmatizing due to their visibility, further deterring consistent user usage. These qualities result in a loss of continuous health monitoring and a critical gap in data collection. In-ear EEG devices offer solutions for these problems by presenting as a discreet, noninvasive method of real-time data collection without interrupting daily routines \cite{kaongoen_future_2023, xu_-ear_2023}. 

\indent This study presents a novel solution to current limitations in the market for EEG recording devices. The development of a personalized in-ear EEG monitor with integrated sound stimulus capabilities is useful for capturing and understanding brain states over extended periods in participants’ everyday environments. With personalization as a primary focus, the device employs custom molds to ensure a precise fit for each user. This facilitates accurate electrode placement and consistent contact with the ear canal and other critical regions, thereby enhancing both comfort and data fidelity. The device is specifically designed for long-term wear to support a more sustainable and practical workflow. Additionally, the built-in sound delivery feature offers potential for future adaptation of the device as a consumer-grade earphone.

\begin{figure} 
\centerline{\includegraphics[width=1\columnwidth]{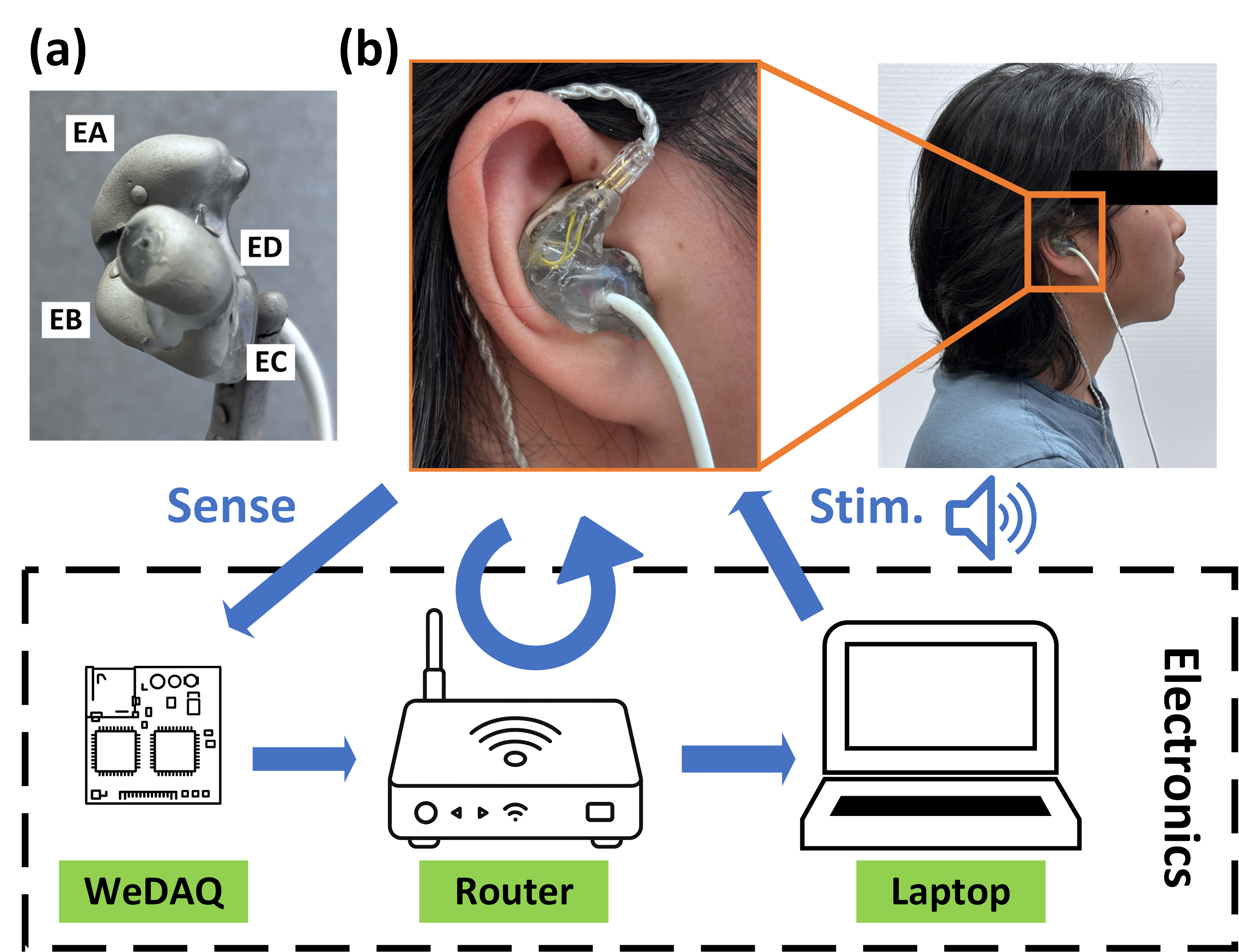}}
\caption{(a) Hardware and on-body setup. Top half shows the on-body picture of the earpiece on the human. Bottom half shows the hardware setup and how it interfaces with the earpiece. (b) IEEM's electrode nomenclature. EA is concha cymba electrode. EB is concha cavum electrode. EC is tragus electrode. ED is ear canal electrode.}
\label{fig:f1}
\end{figure}


\indent There is significant value in developing generic earpieces, as they are inherently more scalable and cost-effective. Prior research has demonstrated the potential benefits of using non–user-specific ear electrode designs \cite{paul_versatile_2022,paul_versatile_2023,lee_scalable_2023}. However, such designs often result in reduced data quality. In many cases, a substantial portion of collected data must be discarded due to electrodes failing to make proper contact and thus not returning usable signals. The anatomy of the ear canal is highly individualized, meaning that taking a generic approach results in loss of reliability. Creating a personalized earpiece based on the unique impression of a user’s ear enhances electrode contact area, while also improving comfort and promoting seamless integration into the user’s daily routine. Furthermore, personalization enables more precise electrode placement in regions known to yield clearer brainwave signals. By conforming to the natural anatomy of the ear, this work aims to appeal to users, researchers, and medical professionals seeking high-fidelity, long-term brainwave monitoring and analysis.

\indent Ear devices or hearables allow for constant monitoring of physiological signals such as brain waves. These devices allow for advances such as the monitoring of signals attributed to both physical and mental conditions \cite{mandic_your_2023}. Electrophysiological signals from the brain have demonstrated value in real-world applications such as diagnosing and preventing diseases and neurological diseases \cite{kamarajan_advances_2015}. 

Furthermore, the integration of sound delivery and EEG recording within the same earpiece enables the development of closed-loop neurofeedback systems. These systems can monitor brain activity—such as alpha rhythms or slow-wave oscillations—in real time and deliver personalized acoustic stimulation accordingly. Recent research has demonstrated the feasibility of using in-ear EEG to enhance sleep quality through auditory neurofeedback, where sound stimuli are modulated based on ongoing EEG dynamics during deep sleep. This closed-loop approach improves sleep architecture by targeting slow-wave activity in a phase-locked manner. The minimally obtrusive and wearable nature of the earpiece allows for practical, long-term use in naturalistic settings
\cite{lee_-ear_2025}. Similarly, other neurofeedback systems have applied real-time EEG-informed interventions to enhance mindfulness meditation by targeting specific brain networks and modulating alpha and theta activity, further highlighting the potential of EEG-driven feedback to support mental well-being \cite{chen_closed-loop_2025}.

\section{Methods and Materials}

\subsection{User-Specific Earpiece Design and Assembly}

\subsubsection{Ear Impression} Our user-specific in-ear EEG monitors (IEEMs) design begins from getting the binaural negative impressions of the ear canal of the subject. The impression is made using a silicone ear mold kit that mixes two-part putty. The otoblock is initially placed just past the first bend in the ear canal to protect the eardrum and ensures the mold does not go too deep. The subject is asked to bite on the bite block to keep the mouth open to further expand the ear canal. Then the putty is stuffed into the outer ear using a syringe immediately after the two parts have been thoroughly mixed, before being cured and still pliable. The putty should be snugged tightly to the curves and grooves to avoid air bubbles, fill both the ear canal and the concha, and remain flush with the outer bowl of the ear.

Once the mold is fully cured, the negative mold impression of the ear is transformed into a computer-aided design (CAD) model using an optical 3D Scanner (Scan Dimension, VA, USA). The CAD model is then processed using Meshmixer (Autodesk, CA, USA) to create a smooth, modified earpiece shell that retains the subject's ear geometry. The shell is in two parts, the first part matches the subject's ear impression, and the second part is the lid that is flush to the outer ear bowl.

\subsubsection{Physical Ear Shell} After finalizing the design, the earpiece shell CAD model is 3D printed using a resin printer (Photon Mono M7 MAX, ANYCUBIC, Shenzhen, China) using a flexible resin material (Superflex, 3D Materials, Gyeonggi-do, South Korea). Once printer, the model is test fitted to the subject to ensure proper and comfortable fitting. If not, the process of modifying the CAD model, 3D printing, and test fitting is iterated to ensure a comfortable and snug fit. Once an acceptable print is made, holes for the electrodes are drilled using a handheld rotary tool. Total of five holes were drilled, four for electrodes and one for the sound tube that connects the balanced armature to the ear canal. The four electrode locations are concha cymba, concha cavum, tragus, and ear canal. Another drill is made on the lid part for the cable to pass through.

\subsubsection{Electrodes} A dome shaped silver 24 gauge headpin is used as the connector between the outer electrodes to the inner wires. Once the headpins are inserted in the 24 gauge drilled holes, it is then soldered to the 4-core shielding sheathed wire. The headpins should be flush to the outer surface of the shell.

Next, the earpiece must be prepped for electrode spray-coating. The surface of the shell is coated to increase surface contact. The surface is first cleaned with isopropyl alcohol and kimtech wipes to remove any residual dirt or lint. A peelable solder mask is applied to establish boundaries that define and isolate the electrodes, thereby shaping their final form. For spray-coating, CI-4025 Ag/AgCl ink from Nagase Chemtex (Nagase Group, NY, CA) and acetone were mixed in 1:15 ratio. The mixture is thoroughly mixed prior to application. Following spraying, the coating is cured at 80$^{\circ}$C for 30 minutes. This coating and curing cycle is repeated three times. Ag/AgCl was chosen because it forms a stable, reversible redox couple $\text{Ag} + \text{Cl}^- \rightleftharpoons \text{AgCl} + e^-$, enabling efficient and low-noise conversion of ionic to electronic signals with minimal polarization. Its low impedance, electrochemical stability, and biocompatibility make it well-suited for accurate, long-term EEG recording without gels.

\subsubsection{Audio Assembly} The assembly starts with preparing the audio driver speaker. The GQ-30710 Knowles balanced armature driver speaker is connected to a transparent heat-shrink polyolefin tube using UV glue. Balanced armature drivers were chosen for their compact size, power efficiency, and exceptional clarity, enabling more accurate sound reproduction and the seamless integration of multiple drivers within a small earpiece. A 2200 $\Omega$ Knowles acoustic damper was inserted in the tube approximately 1 cm away from the driver and fixed using heat to tighten the heat shrink tube near the damper. The acoustic damper smoothens and shapes the frequency response, reducing resonant peaks, and enabling precise tuning of acoustic performance with complex electrical filtering. The wires to the balanced armature driver were then soldered to a female socket jack for in-ear monitors. The components are then fixed using small amounts of UV glue. After all components are secured, the lid is closed using small amounts of UV glue.


\begin{figure}
\centerline{\includegraphics[width=1\columnwidth]{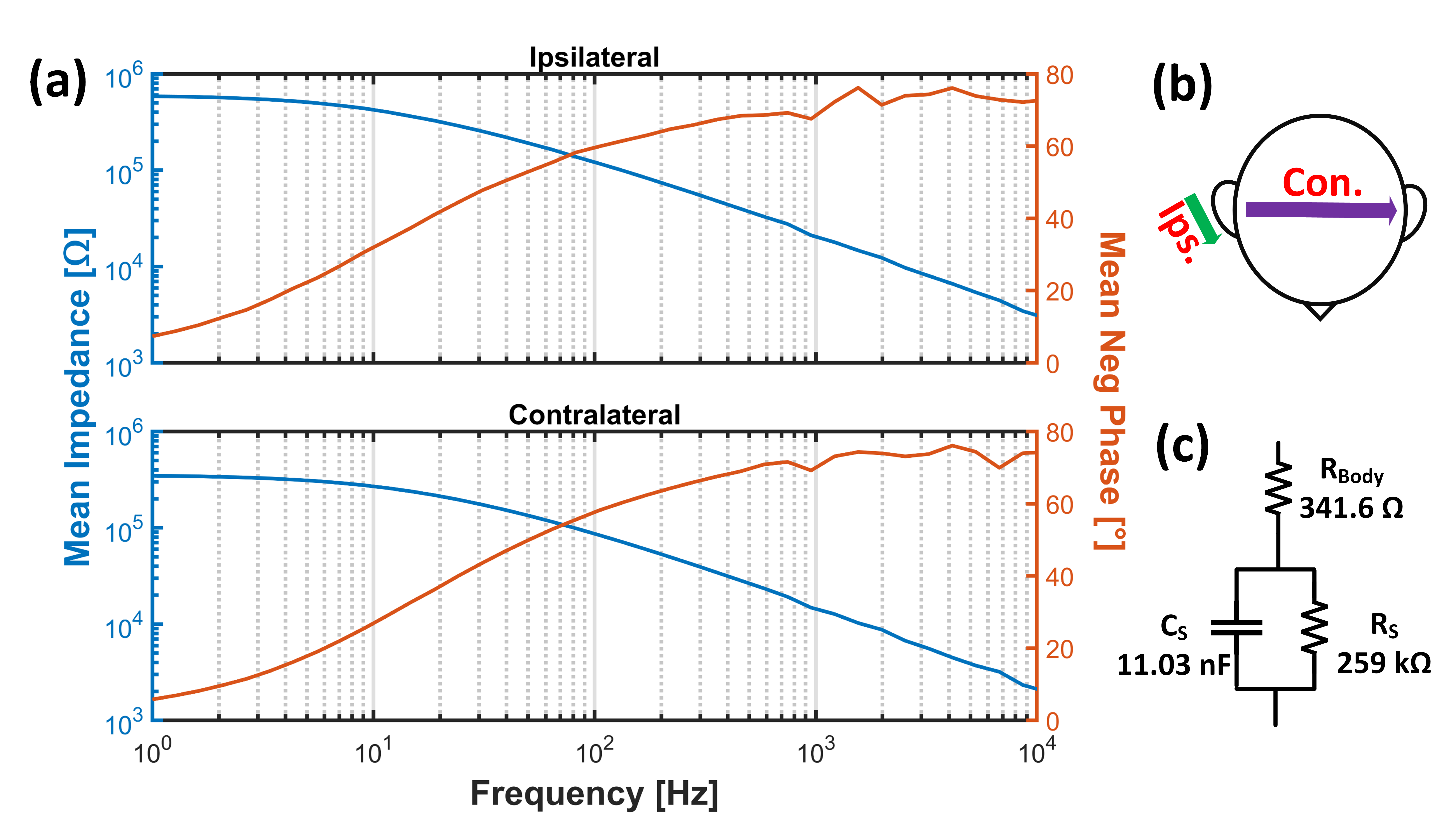}}
\caption{(a) EIS results of both ipsilateral and contralateral configuration. (b) Diagram of the ipsilateral and configuration on human head outline. (c) Randles model of the electrode-skin interface estimated from EIS results.}
\label{fig:f2}
\end{figure}

\subsection{Electrochemical Impedance Spectroscopy (EIS)}

To assess electrode-skin contact quality, a three-electrode electrochemical impedance spectroscopy (EIS) configuration was implemented for each channel while the subject wore the earpiece. Two EIS configurations were employed: ipsilateral and contralateral. In the ipsilateral configuration, the working, counter, and reference electrodes were all placed within the same ear. Specifically, the concha cymba electrode (EA) served as the working electrode, the ear canal electrode (ED) as the counter electrode, and either the concha cavum (EB) or the tragus electrode (EC) was used as the reference electrode in separate recordings (Fig~\ref{fig:f1}). In the contralateral configuration, the working electrode remained in one ear, while the reference and counter electrodes were placed in the opposite ear. The right ear's concha cymba electrode was served as the working, the left concha cymba electrode for the counter electrode, and all other electrodes were used as the reference electrode in separate recordings. EIS measurements were conducted for both configurations in each ear for the ipsilateral configuration, but not the contralateral (Fig. ~\ref{fig:f2}). Measurements were conducted using a PalmSens4 potentiostat with PSTrace software (PalmSens, Utrecht, Houten, Netherlands). The experimental parameters matched those from M. S. Lee et al.: galvanostatic impedance spectroscopy with a pretreatment current range of 100 pA to 100 $\mu$A, applied current of 100 $\mu$A (fixed scan type, i$_{dc}$ = 0.0, i$_{ac}$ = 0.01), and frequency sweeps from 0.1 Hz to 10 kHz \cite{lee_characterization_2022, lee_scalable_2023}.

\subsection{Hardware Setup} 

The hardware configuration for electrophysiological (EXG) measurements is illustrated in Figure~\ref{fig:f2}. In-ear EEG monitoring (IEEM) electrodes are connected to a wireless electrophysiology data acquisition board (weDAQ) using shielded wire cables. The weDAQ board was selected for its low-noise, 24-bit analog-to-digital conversion (ADC) provided by ADS1299 chips, as well as its high-bandwidth wireless data streaming over an 802.11n Wi-Fi protocol using an ESP32 microcontroller. Additional technical specifications of the weDAQ board are detailed in our previous work \cite{paul_versatile_2022, lee_scalable_2023}.

The weDAQ board communicates wirelessly with a remote router via its integrated Wi-Fi protocol, and the router is connected to a laptop over the same wireless network. The laptop runs a custom Python program that interfaces with the weDAQ board through its application programming interface (API), enabling configuration of board settings and real-time data acquisition. Data streams are managed and saved using the Lab Streaming Layer (LSL) protocol.

The weDAQ board was configured with the following settings: a referential (pseudo-monopolar) montage, in which all channels share a common reference electrode; a sampling rate of 500 Hz; a voltage gain of 24; an active driven right leg (DRL) circuit; and continuous impedance monitoring enabled.

For EEG experiments involving auditory stimulation, the earpiece audio driver is connected to the laptop’s audio output jack via a wired connection. The sound stimulus is delivered at an approximate sound pressure level of 50 dBA.

\subsection{Electrophysiology Measurements and Analysis}
Four electrophysiological (EXG) measurements were performed to evaluate and validate the IEEM for on-body biometric measurement: electromyography (EMG), eye blinks, electroculography (EOG), and EEG. Although the core signal is EEG, additional EXG that are likely to couple into the EEG recording were also measured to determine their signal characteristics. The first three measurements were performed to characterize the likely EXG biosignals that will be superimposed on the EEG as artifacts. The characterization of these signals will help better differentiate superimposed signal to understand what artifacts are being coupled. In addition, these signals themselves can be used separate apart from the EEG for their own application. 

\subsubsection{EMG} EMG signals corresponding to jaw clench were recorded from a subject during alternating periods of teeth-gritting and relaxation. Each trial consisted of a 3-second interval during which the subject clenched their jaw, followed by a period of relaxation. This cycle was repeated to achieve a total trial duration of 30 seconds. 
For the time-series plot, a fourth-order Butterworth notch filter centered at 60 Hz was used to remove 60 Hz powerline noise, and a fourth-order Butterworth bandpass filter with cutoff frequencies of 2 and 40 Hz was used. For the spectrogram, a second-order Butterworth notch filter centered at 60 Hz was applied to remove the 60 Hz powerline noise.
\subsubsection{Eye blinks} Eye blinks were recorded as the subject blinked in time with a 60 beats per minute (BPM) metronome for 30 seconds. To eliminate 60 Hz interference and DC offset, the data were processed using a tenth-order Butterworth notch filter centered at 60 Hz and a fourth-order Butterworth bandpass filter with cutoff frequencies of 0.5 Hz and 30 Hz, respectively. Dashes were used to mark each eye blink synchronized with the 60 BPM metronome cues.
\subsubsection{EOG} EOG was recorded while the subject performed eye movements, shifting their gaze left and right in synchrony with a 60 BPM metronome for 30 seconds. To reduce 60 Hz interference and DC offset, the recorded data were processed using a tenth-order Butterworth notch filter centered at 60 Hz and a fourth-order Butterworth bandpass filter with cutoff frequencies of 0.5 to 30 Hz. Only the contralateral (left and right) EOG signals are presented in the results, as the ipsilateral EOG—along with contralateral up and down movements—did not display noticeable activity. This lack of signal is likely due to the movement polarity not being orthogonal to the electrode vector.

\begin{figure} 
\centerline{\includegraphics[width=1\columnwidth]{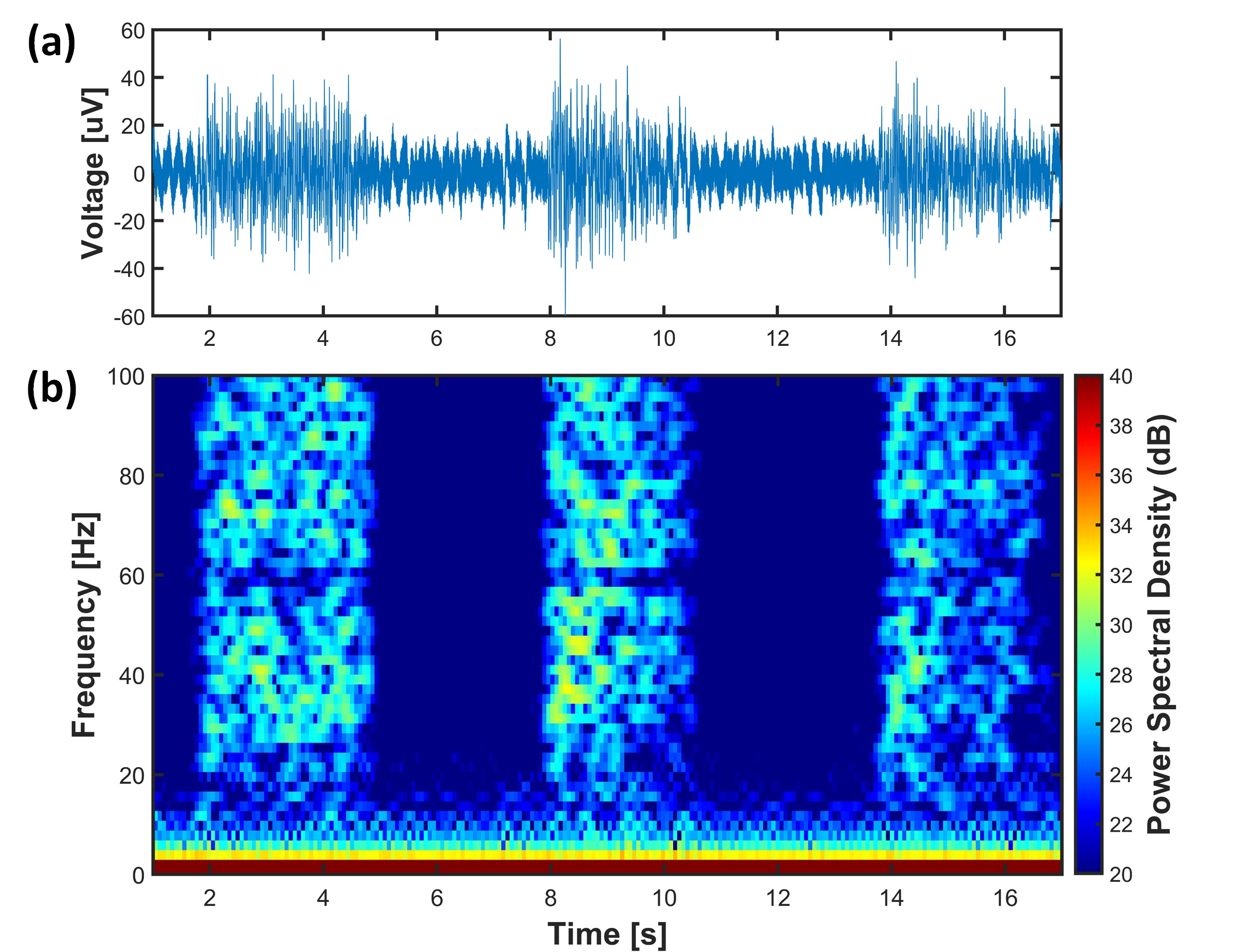}}
\caption{(a) Time-series result of the EMG experiment. (b) Spectrogram result of the same EMG experiment in (a) showing 3 jaw clenches exhibiting broad frequency signal characteristics.}
\label{fig:f3}
\end{figure}

\subsubsection{ASSR and Alpha Modulation} Two EEG modalities were recorded: auditory steady-state response (ASSR) and alpha modulation. ASSR is an EEG-based measurement that captures the brain’s synchronized response to rapidly presented periodic auditory stimuli. This technique is widely used in audiology and neuroscience to objectively estimate hearing thresholds, serving as an alternative to behavioral response methods. Because ASSR does not require active participation from the subject, only the ability to hear the stimulus, it is particularly valuable in consumer and research settings where reliable behavioral responses are unavailable. For our experiment, the ASSR is elicited by having the subject hear a uniform white noise fully amplitude-modulated at 40 Hz at roughly 50 dBA on both ears. 

Alpha modulation refers to dynamic changes in the power of alpha-band oscillations (8–13 Hz) that occur in response to cognitive tasks, attention, or sensory processing. Such modulation is characterized by increases or decreases in alpha power over specific cortical regions. In attention and working memory tasks, increased alpha power is often associated with the suppression of irrelevant or distracting sensory information, while decreased alpha power typically indicates enhanced processing of relevant stimuli. For our experiment, the alpha modulation is elicited by having the subject close their eyes, which is a standard protocol in many EEG studies.

For the complete experiment, the subject kept their eyes closed for the entire duration of the experiment, which was 70 seconds long. The initial five seconds and final 10 seconds were trimmed to remove filter artifacts and any transient effects associated with the onset and offset of the experiment. At the 30-second mark, the 40 Hz amplitude-modulated white noise is played for the rest of the experiment. For visualization of the results, a spectrogram was used with a Hamming window.

\section{Results}

\subsection{Electrochemical Impedance Spectroscopy (EIS)}
In Fig.~\ref{fig:f2}, the mean impedance and the mean negative phase for ipsilateral and contralateral configurations were plotted. The mean impedance at 10 Hz is 424 k$\Omega$ for ipsilateral and 270 k$\Omega$ for contralateral and at 100 Hz is 121 k$\Omega$ for ipsilateral and 87.1 k$\Omega$ for contralateral. For the estimated values of the Randles model for the dry-electrode, the $R_{body}$ is 341.6 $\Omega$, the $C_S$ is 11.03 nF, and the $R_S$ is 2.6 M$\Omega$.

\begin{figure} 
\centerline{\includegraphics[width=0.95\columnwidth]{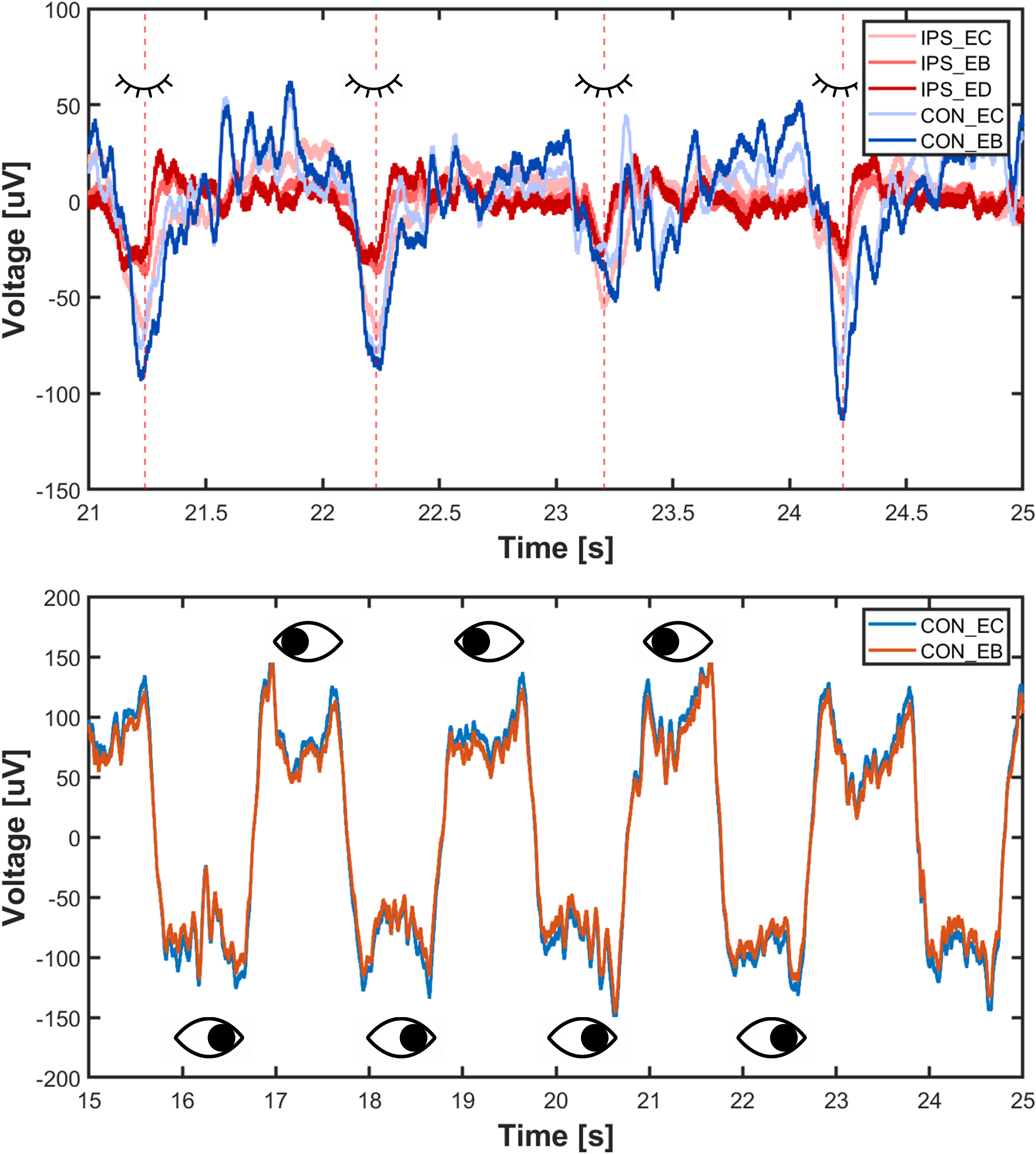}}
\caption{(a) Time-series result of the eye blinks experiment. Red tone color plots are ipsilateral and blue tone color plots are contralateral channels. Eye icon with vertical dash lines were overlayed to indicate when eye blinks occurred. (b) Time-series result of the EOG experiment. Eye icon placed next to signal change to indicate the eye movement direction.}
\label{fig:f4}
\end{figure}

\subsection{Electrophysiology Measurement}
\begin{figure} 
\centerline{\includegraphics[width=1\columnwidth]{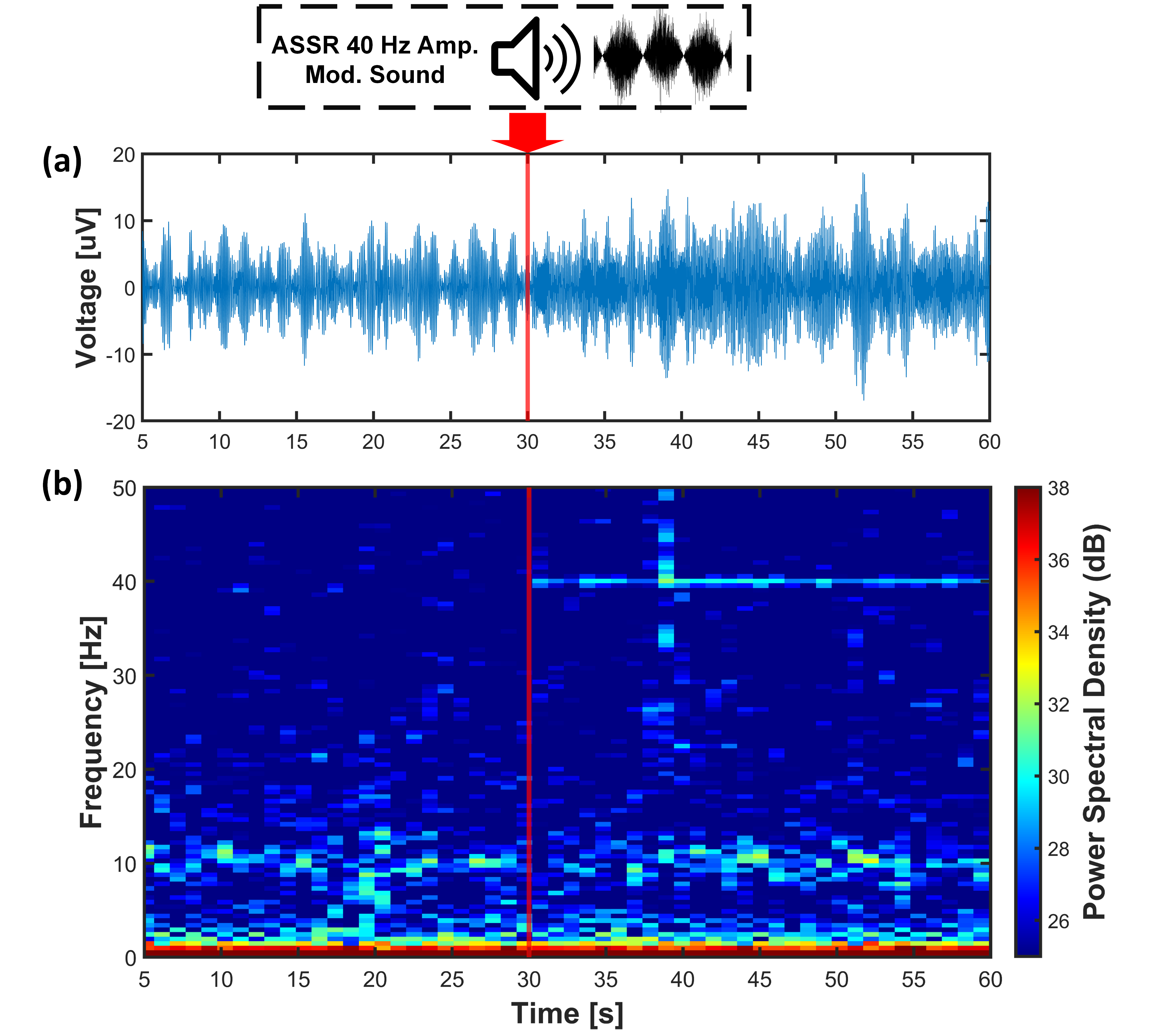}}
\caption{ASSR + alpha experiment with 40 Hz amplitude-modulated white noise sound played at 30s into the experiment. (a) Time-series result of the ASSR + alpha experiment (b) Spectrogram result of the experiment.}
\label{fig:f5}
\end{figure}

\subsubsection{EMG}
In Fig.~\ref{fig:f3}, the spectrogram of the 3-second jaw clench and 3-second relaxation is shown in amplitude dB. The average power at 60 Hz is 2703 $\mu V^2$ before the notch filter. The average power between 1 to 100 Hz besides 60 Hz with $\pm$2 Hz during the relaxation stage is 4.4778 $\mu V^2$. The average power with the same frequency range during jaw clench is 20.5651 $\mu V^2$.

\subsubsection{Eye Blinks}
The eye blinks for 3 ipsilateral and 2 contralateral channels are shown in Fig.~\ref{fig:f4}a. One contralateral channel was omitted due to the noticeable noise. For ipsilateral electrodes, the voltage peaks at the eye blinks are smaller than the contralateral, with the expectation of the ipsilateral electrode located at the ear canal (ED). It may mainly be because of the larger distance between the reference electrode located at the concha cymba (EA) and the ear canal (ED).

\subsubsection{EOG} Fig.~\ref{fig:f4}b shows the EOG of the left and right movement of the eyes. The EOG of the up and down movement of the eyes and ipsilateral configuration is not illustrated due to lack of noticeable change, which may be due to the eyes polarity changes being mainly orthogonal to the electrode configuration. The voltage peak to peak difference is roughly 80 $\mu V^2$ after the digital signal processing.

\subsubsection{ASSR and Alpha Modulation} The Fig.~\ref{fig:f5} shows the spectrogram in part b and a time series plot of the added two filtered signals at 8--13 Hz and 40$\pm$1 Hz in part a. The average power of the alpha-band 8-13 Hz for 5 to 60 seconds is 6.62 $\mu V^2$. The average power of the 40 Hz with $\pm$1 Hz between 30 to 60 seconds is 5.76 $\mu V^2$.

\section{Future Works} 
This work mainly focuses on the initial prototyping and feasibility of this earpiece design. There are several limitations to this work noteworthy to mention. Due to the design being custom and not user-generic, this work only used one subject for the initial validation. In addition, on contralateral channel was showing considerable noise despite having similar impedance as the rest of the electrodes. Further investigation is needed to determine the root cause and see if it relates to the design. In the future, we plan to do additional validations, such as quantifying and distinguishing the superimposed EXG signals, changing the reference and DRL locations, including multi-subject, investigating sound stimulus coupling to EEG signal, and including EEG experiments for functional use.

\section{Conclusion}

This work demonstrates the initial validation of a personalized, custom-molded in-ear EEG monitor that is designed for robust, long-term electrophysiological recording and sound stimulus directly from the ear. By precisely matching each user’s unique anatomy, the device targets stable electrode-skin contact, effective sound isolation, and a flexible softshell for comfortable wear. Although the custom fabrication process limits large-scale deployment, it delivers high-fidelity detection of electrooculography, eye blinks, jaw clenches, auditory steady-state responses, and alpha modulation, making it well-suited for research and wearable health applications.





\section*{Acknowledgment}

The authors thank Alan Brown from Nagase Chemtex for providing a sample and consultation regarding the Ag/AgCl ink. 

\bibliographystyle{IEEEtran}
\bibliography{IEEEabrv, refs_with_doi_2}

@inproceedings{lee_-ear_2025,
	title = {In-{Ear} {EEG} {Auditory} {Neurofeedback} {Towards} {Unobtrusive} {Sleep} {Enhancement}},
	url = {https://ieeexplore.ieee.org/document/10982851/},
	doi = {10.1109/CICC63670.2025.10982851},
	urldate = {2025-05-26},
	booktitle = {2025 {IEEE} {Custom} {Integrated} {Circuits} {Conference} ({CICC})},
	author = {Lee, Min Suk and Liu, Zhaoyi and Uppal, Abhinav and Song, Jiahao and Paul, Akshay and Chapotot, Florian and Tasali, Esra and Xu, Yuchen and Cauwenberghs, Gert},
	month = apr,
	year = {2025},
	note = {ISSN: 2152-3630},
	pages = {1--7},
}

@article{paul_versatile_2023,
	title = {A {Versatile} {In}-{Ear} {Biosensing} {System} and {Body}-{Area} {Network} for {Unobtrusive} {Continuous} {Health} {Monitoring}},
	volume = {17},
	issn = {1932-4545, 1940-9990},
	url = {https://ieeexplore.ieee.org/document/10115033/},
	doi = {10.1109/TBCAS.2023.3272649},
	language = {en},
	number = {3},
	urldate = {2023-11-07},
	journal = {IEEE Transactions on Biomedical Circuits and Systems},
	author = {Paul, Akshay and Lee, Min S. and Xu, Yuchen and Deiss, Stephen R. and Cauwenberghs, Gert},
	month = jun,
	year = {2023},
	pages = {483--494},
}

@inproceedings{paul_versatile_2022,
	address = {Austin, TX, USA},
	title = {A {Versatile} {In}-{Ear} {Biosensing} {System} for {Continuous} {Brain} and {Health} {Monitoring}},
	isbn = {978-1-66548-485-5},
	url = {https://ieeexplore.ieee.org/document/9937994/},
	doi = {10.1109/ISCAS48785.2022.9937994},
	language = {en},
	urldate = {2023-11-07},
	booktitle = {2022 {IEEE} {International} {Symposium} on {Circuits} and {Systems} ({ISCAS})},
	publisher = {IEEE},
	author = {Paul, Akshay and Lee, Min and Xu, Yuchen and Deiss, Stephen and Cauwenberghs, Gert},
	month = may,
	year = {2022},
	pages = {620--624},
}

@article{xu_-ear_2023,
	title = {In-ear integrated sensor array for the continuous monitoring of brain activity and of lactate in sweat},
	volume = {7},
	copyright = {2023 The Author(s)},
	issn = {2157-846X},
	url = {https://www.nature.com/articles/s41551-023-01095-1},
	doi = {10.1038/s41551-023-01095-1},
	language = {en},
	number = {10},
	urldate = {2023-11-03},
	journal = {Nature Biomedical Engineering},
	author = {Xu, Yuchen and De la Paz, Ernesto and Paul, Akshay and Mahato, Kuldeep and Sempionatto, Juliane R. and Tostado, Nicholas and Lee, Min and Hota, Gopabandhu and Lin, Muyang and Uppal, Abhinav and Chen, William and Dua, Srishty and Yin, Lu and Wuerstle, Brian L. and Deiss, Stephen and Mercier, Patrick and Xu, Sheng and Wang, Joseph and Cauwenberghs, Gert},
	month = oct,
	year = {2023},
	note = {Number: 10
Publisher: Nature Publishing Group},
	pages = {1307--1320},
}

@article{lee_characterization_2022,
	title = {Characterization of {Ag}/{AgCl} {Dry} {Electrodes} for {Wearable} {Electrophysiological} {Sensing}},
	issn = {2673-5857},
	journal = {Frontiers in Electronics},
	author = {Lee, Min Suk and Paul, Akshay and Xu, Yuchen and Hairston, W David and Cauwenberghs, Gert},
	year = {2022},
	note = {Publisher: Frontiers},
	pages = {9},
}

@inproceedings{lee_scalable_2023,
	title = {Scalable {Anatomically}-{Tunable} {Fully} {In}-{Ear} {Dry}-{Electrode} {Array} for {User}-{Generic} {Unobtrusive} {Electrophysiology}},
	url = {https://ieeexplore.ieee.org/abstract/document/10340888},
	doi = {10.1109/EMBC40787.2023.10340888},
	urldate = {2024-09-18},
	booktitle = {2023 45th {Annual} {International} {Conference} of the {IEEE} {Engineering} in {Medicine} \& {Biology} {Society} ({EMBC})},
	author = {Lee, Min Suk and Paul, Akshay and Joung, Tae Houn and Xu, Yuchen and Wu, Jiajia and Hairston, W. David and Cauwenberghs, Gert},
	month = jul,
	year = {2023},
	note = {ISSN: 2694-0604},
	pages = {1--4},
}

@article{kaongoen_future_2023,
	title = {The future of wearable {EEG}: a review of ear-{EEG} technology and its applications},
	volume = {20},
	issn = {1741-2552},
	shorttitle = {The future of wearable {EEG}},
	url = {https://dx.doi.org/10.1088/1741-2552/acfcda},
	doi = {10.1088/1741-2552/acfcda},
	language = {en},
	number = {5},
	urldate = {2024-10-30},
	journal = {Journal of Neural Engineering},
	author = {Kaongoen, Netiwit and Choi, Jaehoon and Choi, Jin Woo and Kwon, Haram and Hwang, Chaeeun and Hwang, Guebin and Kim, Byung Hyung and Jo, Sungho},
	month = oct,
	year = {2023},
	note = {Publisher: IOP Publishing},
	pages = {051002},
}

@article{mandic_your_2023,
	title = {In {Your} {Ear}: {A} {Multimodal} {Hearables} {Device} for the {Assessment} of the {State} of {Body} and {Mind}},
	volume = {14},
	issn = {2154-2317},
	shorttitle = {In {Your} {Ear}},
	url = {https://ieeexplore.ieee.org/document/10443768},
	doi = {10.1109/MPULS.2024.3357008},
	number = {6},
	urldate = {2025-05-26},
	journal = {IEEE Pulse},
	author = {Mandic, Danilo and Bermond, Matteo and Occhipinti, Edoardo and Davies, Harry J. and Hammour, Ghena and Nassibi, Amir},
	month = nov,
	year = {2023},
	pages = {17--23},
}

@article{kamarajan_advances_2015,
	title = {Advances in {Electrophysiological} {Research}},
	volume = {37},
	issn = {2168-3492},
	url = {https://www.ncbi.nlm.nih.gov/pmc/articles/PMC4476604/},
	number = {1},
	urldate = {2025-05-26},
	journal = {Alcohol Research : Current Reviews},
	author = {Kamarajan, Chella and Porjesz, Bernice},
	year = {2015},
	pmid = {26259089},
	pmcid = {PMC4476604},
	pages = {53--87},
}

@article{chen_closed-loop_2025,
	series = {Cognitive {Neuroscience} of {Mindfulness}},
	title = {Closed-{Loop} {Systems} and {Real}-{Time} {Neurofeedback} in {Mindfulness} {Meditation} {Research}},
	volume = {10},
	issn = {2451-9022},
	url = {https://www.sciencedirect.com/science/article/pii/S2451902224003112},
	doi = {10.1016/j.bpsc.2024.10.012},
	number = {4},
	urldate = {2025-05-26},
	journal = {Biological Psychiatry: Cognitive Neuroscience and Neuroimaging},
	author = {Chen, Joseph C. C. and Ziegler, David A.},
	month = apr,
	year = {2025},
	pages = {377--383},
}

\end{document}